\documentclass[aps,prb,twocolumn,showpacs,twoside,amsmath, 
amsfonts]{revtex4}
\usepackage{psfrag}
\usepackage{graphicx}
\usepackage{epsfig}
\usepackage{amssymb}

\begin{document}
\title{Two-eigenfunction correlation in a multifractal metal and insulator}
\author{E.Cuevas}
\affiliation{Departamento de F\'{\i}sica, Universidad de Murcia, E30071 Murcia, 
Spain.}
\author{V.E.Kravtsov}
\affiliation{The Abdus Salam International Centre for Theoretical
Physics, P.O.B. 586, 34100 Trieste, Italy, \\ Landau Institute for
Theoretical Physics, 2 Kosygina st., 117940 Moscow, Russia.}

\begin{abstract}
We consider the correlation of two single-particle probability
densities $|\Psi_{E}({\bf r})|^{2}$ at coinciding points ${\bf r
}$ as a function of the energy separation $\omega=|E-E'|$ for
disordered tight-binding lattice models (the Anderson models) and
certain random matrix ensembles. We focus on the
parameter range close but not exactly at the
Anderson localization transition. We show that even away from
the critical point the eigenfunction statistics exhibit the remnant
of multifractality characteristic of the critical states.
This leads to an enhancement of eigenfunction correlations and a corresponding 
enhancement of matrix elements of the local electron interaction at small energy 
separations. This enhancement is accompanied by a depression
of correlations at large energy separations, both phenomena being a consequence of 
the stratification of space into densely packed but mutually avoiding resonance 
clusters. We also demonstrate that the correlation function of 
localized states in a $d$-dimensional insulator is logarithmically enhanced at small
energy separations provided that $d>1$. A simple and 
general physical picture of all these phenomena is presented.
 
Finally by a combination of numerical results on the Anderson model
and analytical and numerical results for the relevant random
matrix theories we identified the Gaussian random
matrix ensembles that describe the multifractal features both in the
metal and in the insulator phases. 
\end{abstract}
 \maketitle
\section{Introduction}
Eigenfunction and spectral statistics in quantum systems with
quenched disorder were a subject of intense study
\cite{Mirlin2000} in the context of mesoscopic fluctuations of
conductance and density of states (DoS), in particular in quantum
dots \cite{ABG}. For this application the most relevant is the
regime of weak deviation \cite{KM,MF} from the Wigner-Dyson
statistics given by the conventional random matrix theory (RMT)
\cite{Mehta}. Disordered multi-channel quantum wires is the most
important example of systems where single-particle eigenstates are
all localized. Here the statistics of eigenstates require a
non-perturbative treatment using the formalism of nonlinear
sigma-model \cite{Efet-book} or banded random matrices
\cite{Mirl-Fyod91}. A special class are systems with the critical,
multifractal (MF) eigenstate statistics  \cite{KMut, CKL,
Mirlin2000}. Two-dimensional disordered metals fall in this class
\cite{EF} provided that effects of localization are suppressed by
magnetic field. Otherwise, one can speak only on weak
multifractality which turns to localization before being fully
developed. The true physical realizations of the critical, MF
eigenstate statistics are systems at the critical point of the
Anderson localization transition \cite{Wegner1980, Ch1990} and the
integer Quantum Hall systems at the center of Landau band
\cite{ChDan}. Importantly, the class of systems with MF eigenstate
statistics also allows for a random matrix representation
\cite{KMut}, in particular using the power-law banded random
matrices (PLBRM) \cite{MFSeil}.

Another field of intense research is the interplay between
disorder and electron interaction with the seminal results on
quantum correction to the tunnel DoS and conductivity
\cite{AAL} of disordered two-dimensional metals and the correction
to superconducting transition temperature due to a simultaneous effect of
disorder and the Coulomb interaction \cite{Finkel}. In all those works
disorder and interaction are taken into account essentially
perturbatively along the lines given in \cite{GLK}. Recently there
was an attempt \cite{FIKY} to consider the problem of
superconductivity near the Anderson transition in which 
disorder has been treated non-perturbatively  by {\it postulating}
the MF statistics of one-particle states $\Psi_{i}(\bf r)$ that
enter the matrix element of a phenomenological electron
attraction:
\begin{equation}
\label{ME} J_{ij} = g\,\int {\bf dr}\, \Psi_{i}({\bf r})^{2}\,
\Psi_{j}({\bf r})^{2}.
\end{equation}
In particular, the simplest  quantity of interest is the disorder
average matrix element $\langle J_{ij} \rangle$ at
a given energy separation $\omega$ between one-particle energies
$E_{i}$ and $E_{j}$. For real eigenfunctions (orthogonal symmetry
class) it is proportional to the correlation function
$C(\omega)=K(\omega)/R(\omega)$, where
\begin{equation}
\label{C} K(\omega) = \int {\bf dr}\,
\sum_{i,j}\langle|\Psi_{i}({\bf r})|^{2}\, |\Psi_{j}({\bf
r})|^{2}\,\delta(E_{i}-E_{j}-\omega)\rangle,
\end{equation}
and $R(\omega)=\sum_{ij}\langle\delta(E_{i}-E_{j}-\omega) \rangle$
is the spectral correlation function which is close to 1 for $\omega$ much greater
then the mean level spacing. 
The correlation function $C(\omega)\approx K(\omega)$ is
the main subject of the present paper.

The correlation function defined by Eq.(\ref{C}) is a measure of
overlap of {\it two different} eigenfunctions. For truly extended
normalized states (e.g. in a quantum dot) $|\Psi_{i}({\bf
r})|^{2}=1/{\cal V}$ and thus $C(\omega){\cal V}=1$, where ${\cal V
}$ is a system volume. Remarkably, $C(\omega){\cal V}=1$ is also
valid for classical examples of localized states, e.g. in a
quantum disordered wire. In this case two states are typically not
overlapping but with a small probability of $\xi^{d}/{\cal V}$
(where $\xi$ is the localization radius) they are localized in the
same place and then the integral in Eq.(\ref{C}) is of the order
of the inverse localization volume $1/\xi^{d}$.

There are cases, however, when an eigenfunction $\Psi_{i}({\bf r})$ does not
occupy all the available volume or all the localization volume 
and the typical amplitude
$|\Psi_{i}({\bf r})|^{2}$ is not just the inverse volume (for
extended states) or the inverse localization volume (for localized
states). In this case a non-trivial behavior of the correlation 
function $C(\omega)$ is expected.
Such situation is realized near the critical point of
the Anderson localization transition. In the vicinity of this
point in the region of extended states ({\it multifractal metal})
or in the region of  localized states ({\it multifractal
insulator}) the system retains the characteristic features of the
critical multifractal statistics of eigenstates which makes it qualitatively
different from both a normal metal or a normal Anderson insulator.

In this paper we will identify and quantify such characteristic
features in the correlation function $C(\omega)$ and give their
interpretation in terms of the typical behavior of single-particle
states. To attain this goal we will combine new analytical results
for the  PLBRM with numerics on the PLBRM and the Anderson model.
We specially focus on the dependence of $C(\omega)$ on the energy
difference $\omega$ in the cross-over region in the vicinity but
not exactly at the Anderson transition point which has not been
studied so far.

The paper is organized as follows. In section II we give a brief
introduction into the subject of 
multifractality of critical eigenstates focusing on  the main effect of
multifractality which is the critical enhancement of eigenfunction
correlations. In section III we give a cartoon of the off-critical states 
in a
multifractal metal and a multifractal insulator and introduce the
random matrix theories which may describe them. In Sec.IV we
present the results of an analytical theory of eigenfunction
correlations for a class of almost diagonal Gaussian random matrices
which all the RMT's suggested to describe strong multifractality fall into.
In Sec.V we consider the two-eigenfunction correlation function
exactly at the critical point of the localization transition in
the 3D Anderson model and for the critical random matrix ensemble
in the limit of strong multifractality. We show that the dynamical
scaling relationship suggested by Chalker is not violated even in
the limit when the fractal dimensions are very small. In Section
VI we describe the new phenomenon of eigenfunction mutual
avoiding and present a qualitative picture that simultaneously
explains the enhancement of eigenfunction correlations at small
energy separations and the eigenfunction mutual avoiding at large
energy separations. In section VII we consider the properties of
eigenfunction correlations in a multifractal insulator. In
particular, we describe the new phenomenon of logarithmic
enhancement of eigenfunction correlations at small energy
separations in the 2D and 3D Anderson insulators and show the
absence of such enhancement in the quasi-1D case. We also
suggest a Truncated Critical RM ensemble that describes all the
principal features of eigenfunction correlations in the 3D
multifractal insulator. Section VIII is devoted to the random matrix
description of the multifractal metal. We show that the
sub-critical PLBRM suggested in Ref.\cite{MFSeil} gives a
reasonable agreement with the 3D Anderson model. By analytical
treatment of this RM model we found the region of parameters where
the eigenfunction correlations become effectively short-range in
the energy space which may poit out on the existence of a new
metal phase above some critical dimensionlity in the
multi-dimensional Anderson model. In the Conclusion we list all
the principal results of this paper.
\section{Multifractality of critical eigenfunctions}
The "standard" model (the Anderson model) for the Anderson
localization transition in $d>2$ dimensions is the tight-binding
model with the hopping constant $V=1$ and random on-site energies
$\varepsilon_{i}$ characterized by the distribution function
${\cal P}(\varepsilon_{i})$ which is frequently chosen constant
${\cal P}(\varepsilon_{i})=1/W$ in the interval $[-W/2, W/2]$ and
zero otherwise. There is a vast literature (see e.g. \cite{Oht}
and references therein) on numerical investigation of the Anderson
localization transition in this model on a 3D lattice. Recently
also higher dimensions $d>3$ become accessible to modern computers
\cite{MEMir}. While the earlier studies of this model were focused on
the critical behavior of the localization/correlation length $\xi$
near the critical disorder $W_{c}$, the recent works were mostly
related with the statistics of critical eigenfunctions. The
multifractality of critical eigenfunctions predicted in
\cite{Wegner1980} almost immediately after emergence of scaling
theory of localization has been confirmed and quantified in
detail.

The results obtained for the Anderson model  exactly at the
critical point seem to be very well described \cite{MirEv} by the
critical PLBRM model \cite{MFSeil, KMut}. This model is defined as
an ensemble of random Hermitean matrices which entries $H_{ij}$
fluctuate independently around zero with the variance:
\begin{equation}
\label{PL} \langle |H_{ij}|^{2}\rangle =
\left\{\begin{matrix}\beta^{-1}, & i=j \cr
\frac{1}{2\left[1+\frac{|i-j|^{2}}{b^2}\right]},& i\neq j\cr
\end{matrix} \right.,
\end{equation}
where $\beta=1,2,4$ for the Dyson orthogonal, unitary, and
symlectic symmetry classes \cite{Mehta} and $b$ is the parameter
that controls the multifractality exponents. This model has been
studied and its comparison with the Anderson model in $d$
dimensions has been done predominantly for the statistical moments
$P_{n}$ of {\it a single} eigenstate at a given energy $E$:
\begin{equation}
\label{IPR} P_{q}(E)=\rho^{-1}\sum_{n}\sum_{{\bf r}}\langle |\Psi_{n}({\bf
r})|^{2q}\,\delta(E-E_{n})\rangle.
\end{equation}
The best known example is the inverse participation ratio (IPR)
given by the second moment $P_{2}$. The multifractal statistics of
a single eigenstate is characterized by the moment $P_{q}$ that
scales with the system volume ${\cal V}$ or the total number of
sites $N$ as:
\begin{equation}
\label{MF} P_{q}\propto \,N^{-(q-1)d_{q}/d},
\end{equation}
where $d_{q}<d$ is the fractal dimension corresponding to $q$-th
moment. The existence of the scaling law Eq.(\ref{MF}) and the
dependence of the exponent $d_{q}$ on $q$ are the principle
features of {\it eigenfunction multifractality}. The fractal
exponents  $d_{q}$ depend also on the symmetry class $\beta$ and
the space dimensionality $d$. For the critical PLBRM Eq.(\ref{PL})
the dependence on $d$ is modeled by the dependence of $d_{q}$ on
the parameter $b$.

The critical scaling Eq.(\ref{MF}) with respect to the system size
$N$ has its {\it dynamical} counterpart when instead of one single
eigenfunction one considers the correlation function  Eq.(\ref{C})
of {\it two} eigenfunctions at an energy separation
$\omega=|E-E'|$ between them. This scaling has been suggested by
Chalker \cite{Ch1990, ChDan} many years ago:
\begin{equation}
\label{ChAnz} C(\omega)=
\frac{1}{N}\,\left(\frac{E_{0}}{\omega}\right)^{\mu},
\;\;\;\;\delta<\omega<E_{0},
\end{equation}
where 
\begin{equation}
\label{d-mu}
\mu=1-\frac{d_{2}}{d},
\end{equation}
$\delta$ is the mean level spacing and $E_{0}$ is the upper
cut-off of multifractality. Numerics on
the integer Quantum Hall systems and in the critical point of the
3D Anderson model was consistent \cite{Huck,SchPot} with this
scaling.

An important feature of Eq.(\ref{ChAnz}) is that the exponent
$1-d_2/d$ in the $\omega$-dependence is smaller than 1. Even in
the limit of infinitely small correlation dimension $d_2$ the
correlation function decays slowly as $1/\omega$. This implies
that the sparse critical states separated by large energy distance
are still well overlaping \cite{FMover}, in contrast to strongly localized
states which typically do not overlap even for nearest neighbors
in the energy space. The reason for such a behavior and the
physical meaning of the energy scale $E_{0}$ will be discussed in
Section VI.

As the correlation function is equal to $C(\omega)\approx 1/N$ both for the
truly  extended and the ideal localized states, Eq.(\ref{ChAnz})
implies the {\it critical enhancement} of eigenfunction
correlations for $\omega < E_{0}$. This enhancement is crucially
important for electron interaction near the Anderson localization
transition, in particular for the superconducting transition
temperature \cite{FIKY}. To illustrate this point we present in Fig.1 
the result of
numerical diagonalization of the critical PLBRM, the classical
Wigner-Dyson RM $\langle |H_{ij}|^{2}\rangle=const$ with extended
eigenstates, and the ensemble of conventional banded random
matrices \cite{Mirl-Fyod91} with exponentially decreasing entries
$\langle |H_{ij}|^{2}\rangle \propto \exp\left(-|i-j|/B\right)$
which describes strongly localized eigenstates in quasi-1D
disordered systems. The critical enhancement of eigenfunction
correlations is evident from this plot.
 \begin{figure}
\includegraphics[width=8cm, height=8cm]{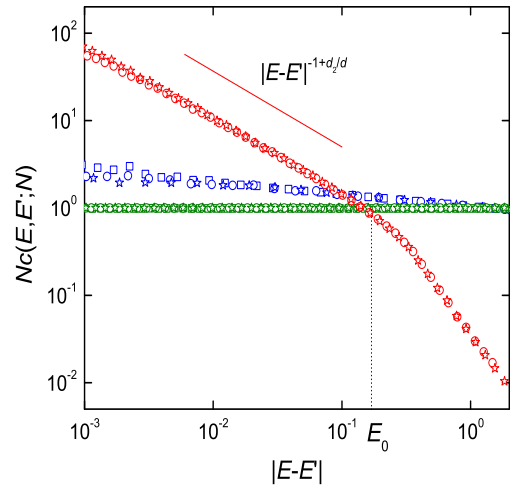}
\caption{Critical enhancement of eigenfunction correlation.
Results of exact diagonalization of the critical PLBRM at $b=0.1$,
the banded random matrices with B=5, and Wigner-Dyson RM are shown
in  red, blue and green, respectively, and are represented by
squares (N=200), circles (N=1000) and stars (N=2000). }
\end{figure}

The physical origin of the enhancement is two-fold: (i) a critical
eigenfunction  "occupies" only part of the available space which
by normalization $\sum_{r}|\Psi({\bf r})|^{2}=1$ enhances its
amplitude, and (ii) the supports (the manifold of $\{{\bf r}\}$
where $|\Psi({\bf r})|^{2}$ is essentially non-zero) of different
critical eigenfunctions are strongly overlapping. It is important
that both conditions are fulfilled simultaneously. For instance
the condition (i) is fulfilled for localized states even better
than for the critical ones but the lack of the condition (ii)
levels off the gain in the correlation function $C(\omega)$. On
the contrary, in a metal the condition (ii) is trivially
fulfilled, but the eigenfunction amplitude is small.

\section{Off-critical states and their random matrix representations}
Gaussian random matrix models proved to be an efficient and
universal theoretical tool for describing complex systems. The
success was partially due to the available analytical solutions
\cite{Mehta, Mirl-Fyod91} and partially due to efficient
algorithms of numerical diagonalization of matrices. Therefore it
is highly desirable to have random matrix models that describe not
only the critical MF eigenstates but also localized and extended
eigenstates in the {\it vicinity} of the Anderson transition. The
criterion to select such models is a qualitative and (when
possible) a quantitative agreement with the results on the 3D
Anderson model.

As will be demonstrated below, the correlation function $C(\omega)$
in the 3D Anderson model contains the critical power-law behavior
Eq.(\ref{ChAnz}) well beyond the Anderson transition point.
As a matter of fact the correlation function $C(\omega)$ is
indistinguishable from the critical one until the dynamic length
$L_{\omega}=1/(\rho \omega)^{1/d}$ ($\rho$ is the mean DoS)
exceeds the localization/correlation length $\xi$. For
$L_{\omega}>\xi$, or $\omega$ smaller than the level spacing in
the localized volume $\delta_{\xi}\sim 1/(\rho \xi^{d})$, the
correlation function  loses its critical features and shows
typical features of a metal or an isulator.

This allows us to suggest the following cartoon of
typical eigenfunctions in the vicinity of the
localization transition shown in Fig.2. 
Namely, a typical localized state in a {\it
"multifractal insulator"} can be viewed as a "piece of multifractal" 
of the size of the  localization radius $\xi$ (Fig.2b.), in contrast to a
conventional localized state where all the
localization volume is more of less homogeneously "filled"
(Fig.2a.). In the same way, typical extended states on the
metallic side of the localization transition ({\it "multifractal
metal"}) should look like a mosaic made of such "pieces of
multifractal" (Fig.2c.).

Based on the persistence of the critical behavior beyond the critical region 
it is 
natural to assume that the random matrix model for the extended states
near the critical point and the localized states on the other side
of the transition  should bear features of the critical RMT. 
\begin{figure}
\vspace{1cm}
\includegraphics[width=0.5cm, height=1cm, angle=270]{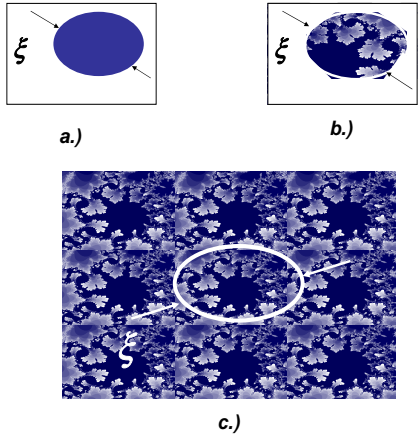}
\vspace{5cm} \caption{2D cartoon of a) conventional localized
state; b) localized state in a multifractal insulator; c) extended
state in a multifractal metal. The darker regions correspond to
higher eigenfunction amplitude. The localization/correlation
radius $\xi$ is shown in each case.}
\end{figure}

Let us start by constructing a random matrix model for the
multifractal insulator. Given that the quasi-1D insulator is
well described by the banded random matrices \cite{Mirl-Fyod91}
with exponentially decaying variance $\langle |H_{ij}|^{2}\rangle
= \exp\{-|i-j|/B\}$  we suggest the following hybrid RM model as a
model for the multifractal insulator:
\begin{equation}
\label{tran} \langle |H_{ij}|^{2}\rangle =
\left\{\begin{matrix}\beta^{-1}, & i=j \cr
\frac{b^{2}}{2\left(|i-j|^{2}+b^{2}\right)}\,\exp\{-(|i-j|/B)^{\eta}\},&
i\neq j\cr
\end{matrix} \right.,
\end{equation}
As
compared with the critical PLBRM model Eq.(\ref{PL}), the model Eq.(\ref{tran}) 
contains an additional parameter $B$ which sets in a
finite localization radius $\xi(B)$. It also contains an exponent
$\eta$ which depends on the space dimensionality $d$ of the
disordered lattice model we would like to model by the RMT. In
Sec.VIII we give both analytical and numerical arguments in favor
of the choice $$\eta=1/d.$$

Another candidate has been suggested in Ref.\cite{MFSeil}:
\begin{equation}
\label{alpha-ins} \langle |H_{ij}|^{2}\rangle =
\left\{\begin{matrix}\beta^{-1}, & i=j \cr
\frac{1}{2\left[1+\left(\frac{|i-j|}{b}\right)^{2\alpha}\right]},&
i\neq j\cr
\end{matrix} \right.
\end{equation}
In this case the localization radius $\xi$ is controlled by the
variable exponent $\alpha$ of the power-law. For a multifractal
insulator $\alpha>1$.

The possible RM models for multifractal metal are also constructed
as deformations of the critical PLBRM. The model
Eq.(\ref{alpha-ins}) for $\alpha<1$ is believed \cite{MFSeil} to
describe the multifractal metal. One can also think that the
Gaussian RMT
\begin{equation}
\label{anti-trank} \langle |H_{ij}|^{2}\rangle =
\left\{\begin{matrix}\beta^{-1}, & i=j \cr
\frac{1}{2\left[1+\frac{|i-j|^{2}}{b^2}\right]}+(b/B)^2,& i\neq
j\cr
\end{matrix} \right.
\end{equation}
which is a hybrid of the critical PLBRM and the WD RMT, is also
suitable for this purpose. Below we will study all those RM models
in detail and compare the corresponding results for the
correlation function $C(\omega)$ with the results obtained by
numerical diagonalization of the $d$-dimensional Anderson model.

\section{Almost diagonal Gaussian RMT: analytical results for $C(\omega)$}
The characteristic properties of multifractal statistics of critical and 
off-critical states are best seen when the multifractality is strong. This
is the case where the parameter $b\ll 1$ in Eqs.(\ref{PL},
\ref{tran}, \ref{alpha-ins}, \ref{anti-trank}) is small. On the
other hand, this is exactly the limit where the typical
off-diagonal elements of $H_{ij}$ are small compared to diagonal
ones. Such matrices (referred to as {\it almost diagonal} random
matrices (ADRM)) may possess a non-trivial statistics of
eigenfunctions which justifies their special study \cite{KYev,
YOs}. The idea of analytical treatment of ADRM, first suggested in
Ref.\cite{Levitov} and used in \cite{MirEv} to compute the
correlation dimension $d_{2}$ for the critical PLBRM model
Eq.(\ref{PL}), is similar to the virial expansion in dilute gases.
However, instead of taking into account two-, three- and
multiple-particle collisions, one considers progressively
increasing number of interacting resonance sites coupled by a small
off-diagonal matrix element  $H_{ij}$. Recently the virial
coefficients for the Gaussian ADRM with an arbitrary (but small)
variance $\sigma^{2}(|i-j|)=2\langle |H_{ij}|^{2}\rangle$ were
expressed through the supersymmetric field theory \cite{YOs} and
the correlation function $C(\omega)$ has been explicitly
calculated in the two-state approximation for the {\it unitary}
symmetry class $\beta=2$. The result is the following:
\begin{equation}
\label{C-K} C(\omega)=\frac{k(\omega)}{N\,r(\omega)},
\end{equation}
where
\begin{equation}
\label{k} k(\omega)=\sum_{n=1}^{N}\left[\left(2\bar{\omega}+
\frac{1}{\bar{\omega}}
\right)\,e^{-\bar{\omega}^{2}}\,\frac{\sqrt{\pi}}{2} \,{\rm
Erfi}\left(\bar{\omega}\right)-1\right],
\end{equation}
and
\begin{equation}
\label{r} r(\omega)=\frac{\sqrt{\pi}}{N}\sum_{n=1}^{N}\bar{\omega}
\,e^{-\bar{\omega}^{2}}\,{\rm Erfi}\left(\bar{\omega}\right).
\end{equation}
In Eqs.(\ref{k},\ref{r}) we denote
\begin{equation}
\label{barO}
\bar{\omega}=\frac{\omega}{\sqrt{2\sigma^{2}(n)}}=\frac{\omega}{2\sqrt{\langle
|H_{i,i+n}|^{2}\rangle }}.
\end{equation}
and ${\rm
Erfi}(z)=\frac{2}{\sqrt{\pi}}\,\int_{0}^{z}e^{t^{2}}\,dt$.

The result given by Eqs.(\ref{k},\ref{r}) is valid in the limit
when:
\begin{equation}
\label{cond} \sum_{n=1}^{N}\sigma(n) \ll \sqrt{\langle
|H_{ii}|^{2} \rangle}.
\end{equation}
For the RMT defined by Eq.(\ref{tran}) and Eq.(\ref{alpha-ins})
with $\alpha>1$ which are suggested to describe the multifractal
insulator, the sum over $n$ in Eq.(\ref{cond})converges. Then the
validity of Eqs.(\ref{k},\ref{r}) is independent of the matrix
size $N$ in the limit $N\rightarrow\infty$ and is controlled only
by a small parameter $b\ll 1$. On the contrary, for the models of
the multifractal metal described by Eq.(\ref{anti-trank}) and
Eq.(\ref{alpha-ins}) with $\alpha<1$ the sum in Eq.(\ref{cond})
diverges at large $N$. Then Eqs.(\ref{k},\ref{r}) are only valid
for $N < \xi$ where we define the {\it correlation radius} $\xi$
as follows:
\begin{equation}
\label{xi} \sum_{n=1}^{\xi}\sigma(n) = \sqrt{\langle |H_{ii}|^{2}
\rangle}.
\end{equation}
We will show below that a good qualitative description of the metal phase 
in the 
limit
$N\rightarrow\infty$ can still be obtained from the above theory
if one substitutes $\xi$ for $N$ in Eqs.(\ref{k},\ref{r}).

Eqs.(\ref{k})-(\ref{xi}) will be used throughout the paper to analize different 
random 
matrix ensembles suggested as possible models for critical eigenstates 
and the off-critical states in a multifractal metal and insulator.

\section{Two eigenfunction correlations at criticality} 
It is not a priori clear that the critical power-law behavior
Eq.(\ref{ChAnz}) and the dynamical scaling relationship Eq.(\ref{d-mu}) hold
true for all
systems where Eq.(\ref{MF}) is valid. In particular it is
interesting to study the correlation function Eq.(\ref{C}) in the
limit of strong multifractality when $d_{2}\rightarrow 0$. Below we will
derive an {\it analytical} formula for the critical PLBRM in the
limit $b\rightarrow 0$ which corresponds \cite{MEMir} to
$d_{2}\rightarrow 0$ and confirm the scaling law Eq.(\ref{d-mu})
by numerical diagonalization of PLBRM with very small $b$.

One can easily see from
Eqs.(\ref{k})-(\ref{barO}) in which we plug in Eq.(\ref{PL}) that in the interval 
$\frac{b}{N}\ll
|E-E'|\ll b$ the correlation function $C(\omega)$ given by these
equations has an asymptotic  power law behavior Eq.(\ref{ChAnz})
with $\mu=1$:
\begin{equation}
\label{crit-b-0}
NC(\omega)=\frac{E_{0}}{|\omega|},\;\;\;\;E_{0}=\left(\frac{\pi}{2}\right)^{3/2}\,b.
\end{equation}

Applying Eq.(\ref{cond}) to the critical PLBRM Eq.(\ref{PL}) gives
the criterion of validity $\ln N\ll 1/b \sim 1/d_{2}$ in which case one cannot
distinguish between $N^{1-d_{2}}$ and $N$, or,
correspondingly, between $\omega^{-1+d_2}$ and $1/\omega$.
Thus the analytical formulae Eqs.(\ref{k},\ref{PL}) is consistent
with the scaling relationship Eq.(\ref{d-mu}), given that $d_{2}(b)\rightarrow 0$
as $b\rightarrow 0$.

A comparison of the analytical results and the results of numerical diagonalization
of the critical PLBRM with $\beta=2$ and $b=0.06$ is shown in
Fig.3. The coincidence is very good for large energy separations. The deviation
at small energy separations is due to the difference in the values of $\mu$.
The exponent $\mu=1$ for the analytical curve and $\mu=0.86$ for the numerical curve
which is very close to the prediction of Eq.(\ref{d-mu}) $\mu=1-d_{2}\approx 0.865$, 
where $d_{2}$
is found from the numerical data for $P_{2}(N)$ and Eq.(\ref{MF}).

The scaling relationship Eq.(\ref{d-mu}) is further checked 
in Fig.4 where the numerical data
for $\mu$ and $1-d_2$ is plotted as a function of $b$. The fulfillment of this
relationship down to $b$ as small as 0.005 and an agreement with the theoretical 
prediction of Ref.\cite{MirEv} is spectacular.
\begin{figure}
\includegraphics[width=9cm, height=9cm]{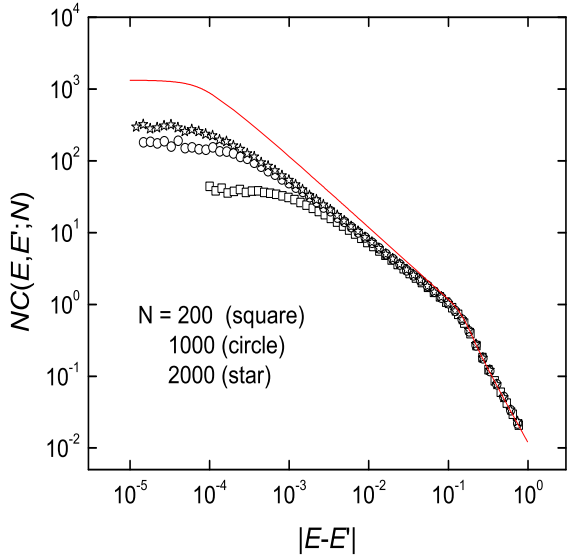}
 \caption{Two-eigenfunction correlation function $C(\omega)$ for the critical PLBRM with
 $\beta=2$,
 $b=0.06$ and $N=200$(square),1000(circle) and 2000 (star). The analytical curve 
at $N=2000$ given by
 Eqs.(\ref{C-K}-\ref{r}) is shown by a solid line.}
\end{figure}
\begin{figure}
\includegraphics[width=8cm, height=8cm]{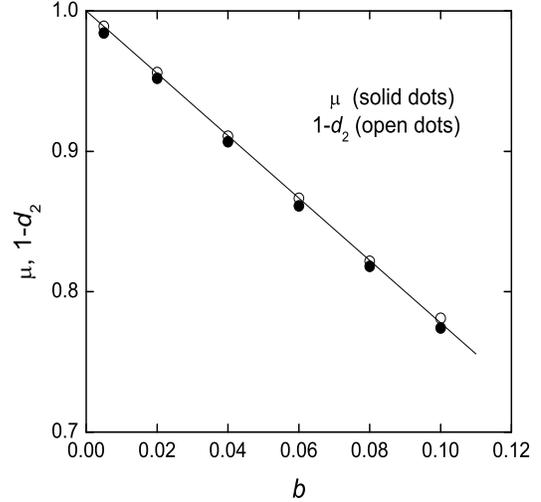}
\caption{The scaling relationship between $\mu$ and $d_{2}$ for
the $\beta=2$ critical PLBRM. The solid line is the prediction
based on Ref.\cite{MirEv} $d_{2}=\frac{\pi}{\sqrt{2}}b$.}
\end{figure}
\begin{figure}
\includegraphics[width=9cm, height=9cm]{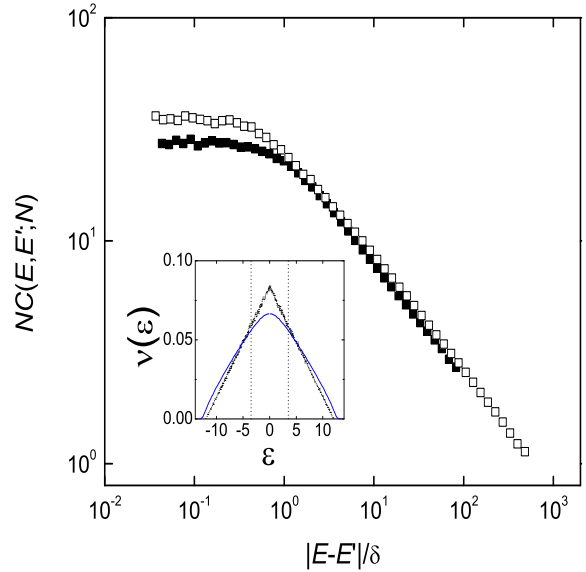}
\caption{Two-eigenfunction correlation function for the 3D
Anderson model (orthogonal symmetry class) with a triangular
distribution of random on-site energies (solid symbols) and the
critical PLBRM Eq.(\ref{PL}) with $\beta=1$ and $b=0.42$ (open
symbols). The energy difference $\omega=|E-E'|$ is measured in
units of mean level spacing. The insert shows the mean density of
states; the mobility edge corresponds to $\varepsilon=\pm 3.5$.
The energies $E,E'$ were taken from the window $(3.3,3.7)$ for the
3D Anderson model and $(-0.2,0.2)$ for the critical PLBRM. The
slope of the critical power-law Eq.(\ref{ChAnz}) is 0.52 in both
cases which corresponds to $d_{2}/d=0.48$.}
\end{figure}
Thus from the combination of analytical and numerical results we
conclude that the Chalker's scaling Eqs.(\ref{ChAnz}),(\ref{d-mu}) is valid for
arbitrary small $b$ and thus for arbitrary strong multifractality.

Finally we demonstrate how well the critical PLBRM Eq.(\ref{PL}) describes the 
two-eigenfunction correlations in the 3D Anderson model at the mobility edge.
To this end we modify the distribution of the on-site energies in the 
Anderson model from the standard rectangular box distribution to the triangular 
distribution where the mobility edge corresponds to $E_{c}=\pm 3.5$.
The correlation function $C(\omega)$ with $E,E'$ near the mobility edge 
is shown in Fig.5. It coincides almost exactly with the corresponding curve resulting 
from
numerical diagonalization of the critical PLBRM ensemble with only one fitting 
parameter $b=0.42$.

\section{Eigefunction mutual avoiding and stratification of coordinate space}
Results of both numerical and analytical calculation presented in
Fig.3 reveal another unexpected feature of eigenfunction
correlation which appears to be common to all ADRM. Surprisingly
it is also present for the 3D Anderson model both in the metal and in 
the insulator phase (see Fig.6). This is
the {\it negative} eigenfuncton correlations for
$\omega=|E-E'|>E_{0}\sim b$. Indeed, one can see from Fig.3 and
Fig.6 that for large enough $\omega$ the correlation function
$C(\omega)$ goes below the uncorrelated limit $C(\omega)=1/N$
which corresponds to
$\langle|\Psi_{i}|^{2}\,|\Psi_{j}|^{2}\rangle=
\langle|\Psi_{i}|^{2}\rangle\,\langle|\Psi_{j}|^{2}\rangle=1/N^{2}$.
We denote by $E_{0}$ the value of $\omega$ where this limit is
reached. For $\omega>E_{0}$ the correlation function
$C(\omega)\propto 1/\omega^{2}$ decreases down to zero.
\begin{figure}
\includegraphics[width=8cm, height=8cm]{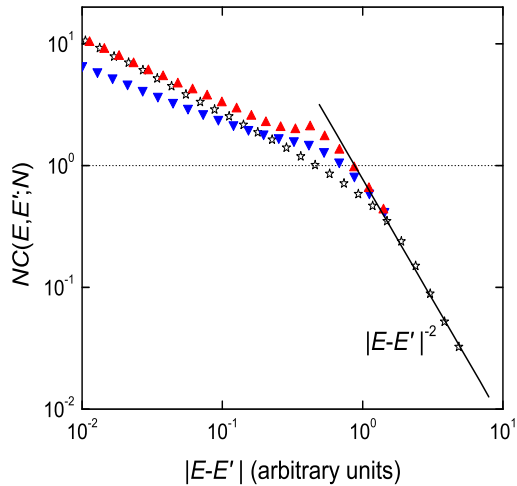}
\caption{Eigenfunction avoiding for the PLBRM  with b=0.42(stars),
3D Anderson insulator (red triangles) and metal (blue triangles).
The dotted line corresponds to the limit of uncorrelated
eigenfunctions; the solid line corresponds to the power law
$1/\omega^{2}$. Points below the dotted line correspond to
eigenfunction avoiding. }
\end{figure}
Such a behavior implies that two eigenfunctions separated by an
energy difference $\omega>E_{0}$ try to {\it avoid each other}.
That is, if a site ${\bf r}$ is occupied in one of the states it
should be predominantly empty in the other.

To explain such a behavior the following cartoon is useful. Let us
define a {\it support} of an eigenfunction as the manifold of
sites $\{ {\bf r}\}$ where $|\Psi_{i}({\bf r})|^{2}$ is
essentially non-zero. To construct such a support starting from a
given site ${\bf r}$ with the on-site energy $\varepsilon_{{\bf
r}}$ we find all the sites  in resonance with the site ${\bf r}$,
i.e. such sites ${\bf r'}$ which on-site energies
$\varepsilon_{{\bf r}'}$ obey the relationship $|\varepsilon_{{\bf
r}'}-\varepsilon_{{\bf r}}|< |H_{{\bf r},{\bf r}'}|$. Then the
procedure should be repeated for all sites ${\bf r}'$ and so on.
It is important that so obtained manifold  $\{{\bf r}\}$ does not
always include all the sites of the system. If this is the case,
the whole coordinate space is {\it stratified} into a set of
mutually non-intersecting supports (Fig.7).
\begin{figure}
\vspace{1cm}
\includegraphics[width=4cm, height=4cm]{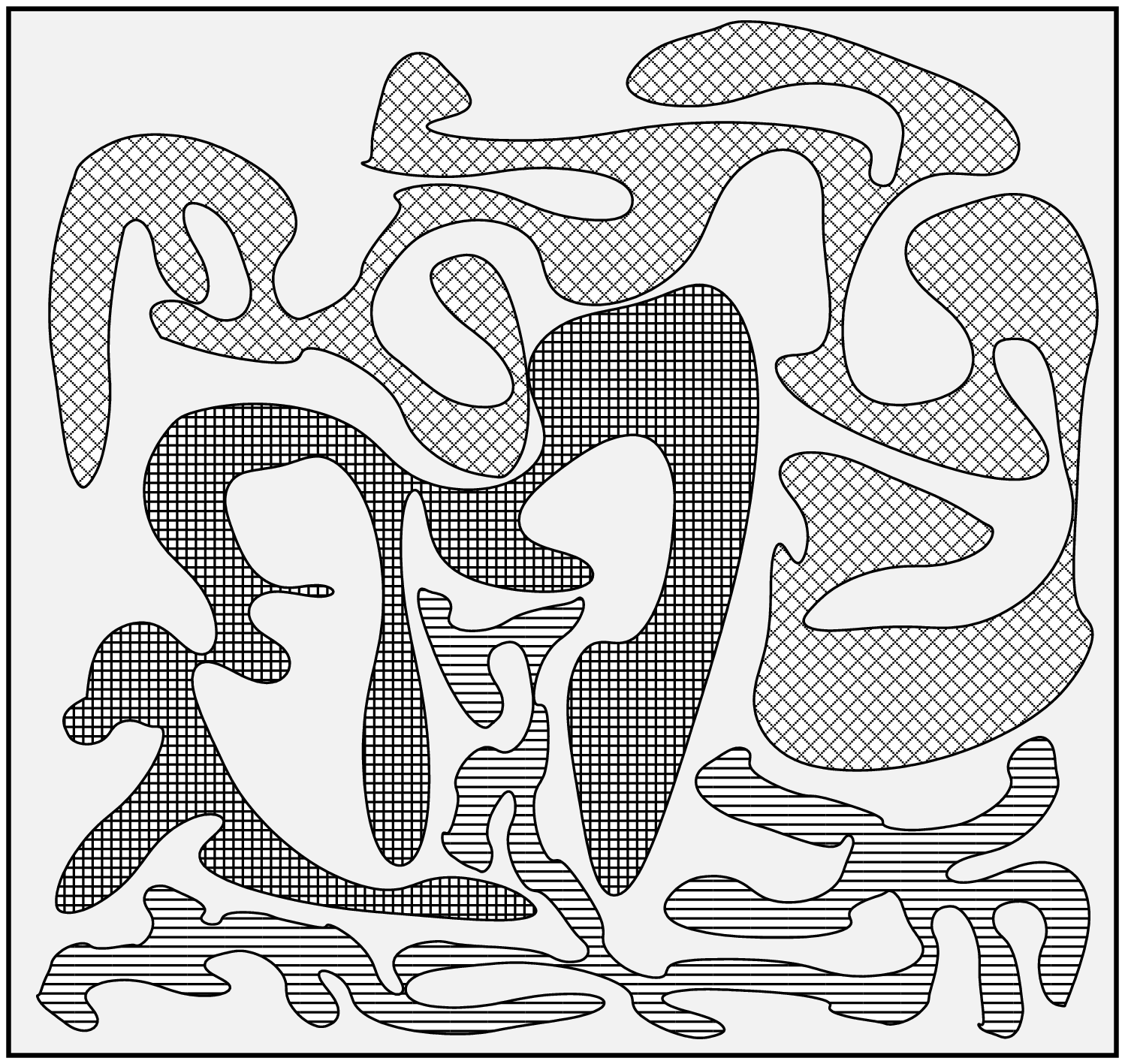}
\caption{A cartoon of stratification of the coordinate space:
different non-intersecting supports shown by different colours.
Each support corresponds to a shell of states occupying this
support and thus strongly overlapping; states belonging to
different shells do not overlap. The stratification of space
explains both strong correlations of states at  energy separation
$\omega$ smaller that the single-shell bandwidth $E_{0}$ and
mutual avoiding of eigenstates for $\omega>E_{0}$.}
\end{figure}

Once the support is defined, one can build a {\it shell} of states
on this support by making a linear combination of on-site states,
pretty much in the same way as in building the conduction band
states out of the on-site states in the tight-binding model. Then
by construction the eigenfunctions belonging to the same shell are
well overlapping but those belonging to different shells do not
overlap.

From this cartoon it is clear that the physical meaning of the
scale $E_{0}$ is the width of the energy band corresponding to a
single shell. Indeed, if the energy separation $\omega$ greatly
exceeds the typical single shell bandwidth, the two eigenfunctions
must belong to different shells and thus do not significantly
overlap in space. On the contrary, if $\omega$ is smaller than the
single shell bandwidth, the two states typically belong to the
same shell and thus overlap strongly no matter how sparse the
shell support.

The new energy scale $E_{0}$, which is the upper energy cut-off of
the multifractal correlations, corresponds to a new length scale
\begin{equation}
\label{ell} \ell_{0}=\frac{1}{(\rho E_{0})^{1/d}},
\end{equation}
which has a meaning of  the minimum length scale of the fractal
texture. In the $d$-dimensional Anderson model the energy scale
$E_{0}$ can be estimated as $$E_{0}\sim V \sim  D/W_{c}\sim
D/(2d\ln2d),$$ where $D$ is the total bandwidth.
Estimating the DOS as $\rho=1/(a^{3}D)$ we find $\ell_{0}\sim a\,
W_{c}^{1/d}$, where $a$ is the lattice constant.

Clearly the  picture with a stratified coordinate space is
possible for PLBRM Eq.(\ref{PL}) with small enough $b<1$ when the
single shell bandwidth $E_{0}\sim b$ is small compared to the
total bandwidth $\sim 1$. Amazingly, the 3D Anderson model which
low-frequency critical features are well described by the critical
PLBRM with $b\approx 0.42$, also follows the predictions of the
critical PLBRM for high frequencies $\omega>E_{0}$. This is a
consequence of a relatively large value $W_{c}=16.5$ of the
critical disorder which results in $E_{0}$ considerably smaller than the
conduction bandwidth $D$. In particular its coordinate space must
be stratified to explain the observed (see Fig.6) mutual avoiding
of eigenstates.

\section{The multifractal insulator}
\begin{figure}
\vspace{1cm}
\includegraphics[width=8cm, height=8cm]{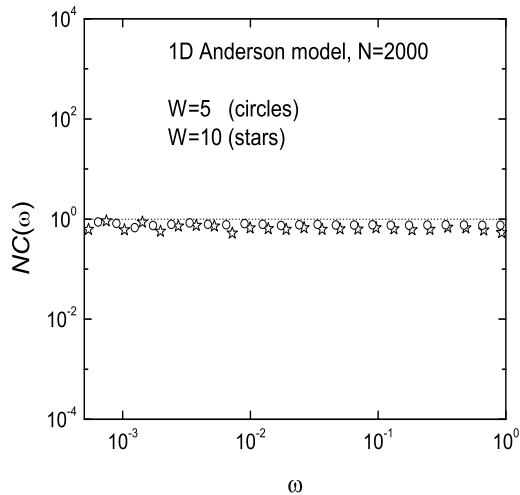}
\caption{Eigenfunction correlation in the 1D Anderson insulator
with rectangular distribution of on-site energies and periodic
boundary conditions. The disorder strength is $W=5$ (circles),
$W=10$ (stars). The inverse participation ratio is equal to 0.23
and 0.46, respectively.}
\end{figure}
As has been demonstrated in Sec.V critical eigenfunctions with
$\xi>L$ are strongly correlated in space. Here we consider the
case of multifractal insulator where the localization radius $\xi$
is large compared to relevant microscopic lengths (the lattice
constant or elastic scattering length) but is much smaller than
the system size $L$. We will identify a suitable random matrix
model to describe this case and compare the properties of
eigenfunction correlation in this model with those of the 3D
Anderson model.
\subsection{The ideal insulator limit}

We start by considering a limit of strong disorder when the
localization length $\xi\sim 1$ and the multifractal nature of
eigenstates does not show up. A common wisdom is that in the
strongly localized regime the positions of the localization
centers  are completely uncorrelated. As it is shown in the
Introduction, this leads to
\begin{equation}
\label{st-loc-l} NC(\omega)=1,
\end{equation}
which we will refer to as the {\it ideal insulator limit}. Fig.1
shows how this limit is reached in the ensemble of banded random
matrices. 

\begin{figure}
\vspace{1cm}
\includegraphics[width=8cm, height=8cm]{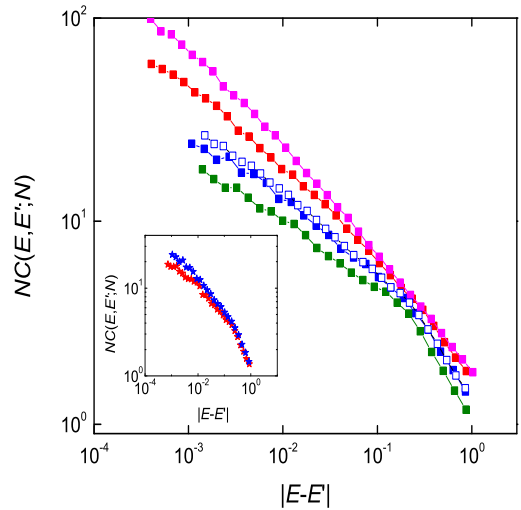}
\caption{Eigenfunction correlation in the 3D Anderson insulator
with rectangular distribution of on-site energies and periodic
boundary conditions. The disorder strength is $W=80$ (green),
$W=60$ (blue), $W=40$ (red), $W=30$ (purple). The system size is
$L=20$ for filled symbols and $L=8$ for open symbols. The inverse
participation ratio for the four insulating systems is
$P_{2}=0.72$, $0.63$, $0.44$, $0.28$ which corresponds to
$\xi=1.0$, $1.1$, $1.2$, $1.4$ according to $\xi=(9/4\pi
P_{2})^{1/3}$. The change of the slope occurs at
$|E-E'|=\delta_{\xi}$. The slope for larger energy separations
$|E-E'|>\delta_{\xi}$ progressively increases with increasing $W$
remaining smaller than 1. The insert shows the result for $W=60$,
$L=20$ for the periodic (upper blue curve) and the hard wall
(lower red curve) boundary conditions.}
\end{figure}
Note that $C(\omega)$in  this limit   is much smaller than  the
self-overlap of $|\Psi({\bf r})|^{2}$ given by the inverse
participation ratio Eq.(\ref{IPR}). Only for very small energy
separations (typically $\propto e^{-L/\xi}$) which we will not be
considering here, the IPR limit can be approached.

Now let us see how does the correlation function $C(\omega)$ look
like for the strong Anderson insulator. The corresponding plot for
1D Anderson model is shown in Fig.8. It  coincides almost exactly
with the ideal insulator limit Eq.(\ref{st-loc-l}).

The plot for the 3D Anderson model is shown in Fig.9. On can see
that $NC(\omega)$ is significantly enhanced at small energy
separations and does not resemble at all the correlation function
in 1D Anderson insulator.
\subsection{Repulsion of centers of localization for $R\gg\xi$}
In order to understand why the ideal insulator limit is not
reached in the 3D case despite the ratio $\xi/L>10$ we compute
numerically the probability distribution (PDF)
\begin{equation}
\label{JPDF} F_{\omega}(R)=\langle
\delta(\omega-E_{n}+E_{m})\,\delta(R-|{\bf r }_{n}-{\bf r}_{m}|)
\rangle
\end{equation}
of the distance $R=|{\bf r}_{n}-{\bf r}_{m}|$ between the points
${\bf r}_{n}$ and ${\bf r}_{m}$ in {\it real space} ({\it centers of
localization}) where $|\Psi_{n,m}({\bf r})|^{2}$ has an absolute
maximum, provided that the  energy separation between the states $n,m$
is $\omega$.

The results are shown in Fig.10. It is seen that the function 
$F_{\omega}(R)$
is far from being independent of $R$ (which would imply the lack
of correlations between centers of localization). In fact, there
is a {\it repulsion of centers of localization} at distances
$R<R_{0}\sim 12$ which shows up in the decreasing probability
density to find two centers of localization close to each other.
Note that $R_{0}$ is almost 10 times larger than the localization
radius $\xi$ estimated from the inverse participation ratio
$P_{2}$.
\begin{figure}
\vspace{1cm}
\includegraphics[width=6cm, height=6cm]{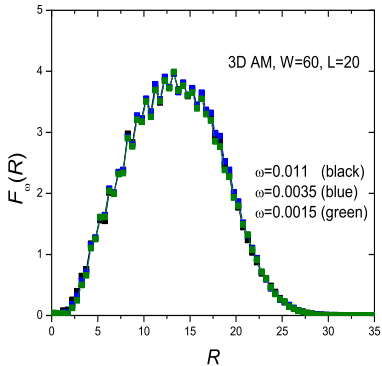}
\caption{The probability density Eq.(\ref{JPDF}) of having two
centers of localization at a distance $R$ in real space and at a
distance $\omega$ in the energy space computed for the 3D Anderson
model in the strong localization regime ($W=60$, $\xi=1.1$,
$L=20$). The repulsive core $R_{0}\sim 10-12$ exceeds the "hard
ball" limit $2\xi$ by a factor of 5-6.}
\end{figure}

An explanation to this fact of repulsion between centers of
localization ${\bf r_{n}}$ and ${\bf r_{m}}$ is based on the
resonance interaction between states $\Psi_{n}({\bf r})$ and
$\Psi_{m}({\bf r})$if the energy distance between them is smaller
than the typical overlap integral $V_{nm}(R)\propto e^{-R/2\xi}$.

The size of the repulsion core $R_{0}$ can be estimated from the
equation:
\begin{equation}
\label{R}V_{nm}(R_{0})=\omega,\;\;\;\Rightarrow R_{0}\sim 2\xi
\,\ln\left(\frac{\delta_{\xi}}{\omega} \right).
\end{equation}
The characteristic energy scale $\delta_{\xi}$ is the mean level
separation for states localized in the same volume $\xi^{d}$. Thus
the repulsion of centers of localization is a direct consequence
of repulsion of energy levels for states confined in the same
volume $\xi^{d}$. The energy scale $\delta_{\xi}$ depends on the
strength of disorder and is of the order of the Fermi energy for
strongly localized states. At $\omega\ll \delta_{\xi}$ the size of
the repulsion core $R_{0}$ may considerably exceed the
localization radius. 

The qualitative picture of repulsion of centers of localization
can be quantitatively confirmed using  the analytical theory
Eqs.(\ref{C-K}-\ref{barO}) for the almost diagonal Gaussian RMT.
To this end we look at the contribution of $\bar{\omega}\gg 1$ to
the sum in Eq.(\ref{k}). Replacing the summation over $n$ by
integration we obtain the contribution to $NC(\omega)$:
\begin{equation}
\label{hvost} \int_{\sigma(n)<
\omega}\frac{2\sigma^{2}(n)}{\omega^{2}}\;dn.
\end{equation}
This equation can be easily interpreted using an elementary
perturbation theory. Indeed, for strongly localized states
$\Psi_{m}({\bf r}_{n})$ the eigenfunction correlation function can
be represented as follows:
\begin{eqnarray}
\label{appr-psi}&& C_{nm}=\sum_{{\bf r}}|\Psi_{n}({\bf
r})|^{2}\,|\Psi_{m}({\bf r})|^{2}\approx\\ \nonumber &\approx&
|\Psi_{m}({\bf r_{n}})|^{2}\sum_{{\bf r}}|\Psi_{n}({\bf
r})|^{2}+|\Psi_{n}({\bf r_{m}})|^{2}\sum_{{\bf r}}|\Psi_{m}({\bf
r})|^{2}\\ \nonumber &\approx&|\Psi_{m}({\bf
r_{n}})|^{2}+|\Psi_{n}({\bf r_{m}})|^{2} .
\end{eqnarray}
The amplitude at the tail of the wavefunction $|\Psi_{n}({\bf
r_{m}})|^{2}$ with the maximum at a point ${\bf r}_{n}$ can be
computed from the elementary perturbation theory in which the
wavefunction of the zero-th approximation corresponding to the
energy $\varepsilon_{n}$ is $|\Psi_{n}^{(0)}({\bf r
})|^{2}=\delta_{{\bf r},{\bf r}_{n}}$:
\begin{equation}
\label{PT} |\Psi_{n}({\bf r_{m}})|^{2}=|\Psi_{m}({\bf
r_{n}})|^{2}\approx\frac{|H_{nm}|^{2}}{(\varepsilon_{n}-\varepsilon_{m})^{2}}
\approx \frac{|H_{nm}|^{2}}{\omega^{2}}\ll 1.
\end{equation}
The fluctuating {\it on-site} energy $\varepsilon_{n}$ is the main
part of the eigenvalue $E_{n}$ for a sufficiently strongly
localized state. Thus we come to a conclusion that the amplitude
of the wavefunction $\Psi_{n}$ at a center of localization of the
wavefunction $\Psi_{m}$ is inversely proportional to
$(E_{n}-E_{m})^{2}=\omega^{2}$ and thus is strongly {\it enhanced} when
$\omega\ll \delta_{\xi}$. At the first 
glance this is in a
contradiction with the common wisdom that $|\Psi_{n}({\bf
r}_{m})|^{2}\propto e^{|{\bf r}_{n}-{\bf r}_{m}|/\xi}$ which is
apparently $\omega$-independent. The point is that the quantity
$|\Psi_{n}({\bf r})|^{2}$ 
has many accidental spikes due to resonances between on-site energies.
The measure of such resonance points is small and for some (but not all)
purposes one can neglect them and to approximate $|\Psi_{n}({\bf r})|^{2}\propto
e^{|{\bf r}_{n}-{\bf r}_{m}|/\xi}$. The best known example when the two-spike
eigenfunction makes the main contribution is the low-frequency conductivity
in the localized phase \cite{Mott}. As we will see below, here we deal with 
a very similar phenomenon.

Now the correlation function $C(\omega)$ can be computed just by
averaging over disorder and the distance ${\bf R}={\bf r}_{n}-{\bf
r}_{m}$:
\begin{eqnarray}
\label{C-finger} C(\omega)&=&\int F_{\omega}(R)\,\langle
C_{nm}(R)\rangle\,d^{d}R \\ \nonumber &\approx& \int
F_{\omega}(R)\,\frac{2\langle
|H_{nm}|^{2}\rangle}{\omega^{2}}\,d^{d}R,
\end{eqnarray}
where $F_{\omega}(R)$ is the PDF defined by Eq.(\ref{JPDF}).

Comparing Eq.(\ref{C-finger}) for $d=1$ with Eq.(\ref{hvost}) we
see that:
\begin{equation}
\label{F-w} F_{\omega}(R)=1/N,\;\;\;\; (R\gg R_{0}),
\end{equation}
where $R_{0}$ is found from the condition $\sigma(R_{0})=\omega$
similar to Eq.(\ref{R}).

In the opposite limit $R\ll R_{0}$, or $\omega\ll \sigma(R)$, we
have a resonance enhancement $|\Psi_{m}({\bf r}_{n})|^{2}\approx
1/2$ and $C_{nm}\approx 1$. Then the comparison of
Eq.(\ref{C-finger}) with Eq.(\ref{k}) yields:
\begin{equation}
\label{p-gap} F_{\omega}(R)=
\frac{4}{3N}\,\left(\frac{\omega}{\sigma(R)} \right)^{2},
\;\;\;\;(R \ll R_{0}).
\end{equation}
\subsection{Logarithmic enhancement of correlations of localized eigenfunctions and the
Truncated Critical RMT}
\begin{figure}
\vspace{1cm}
\includegraphics[width=7cm, 
height=7cm]{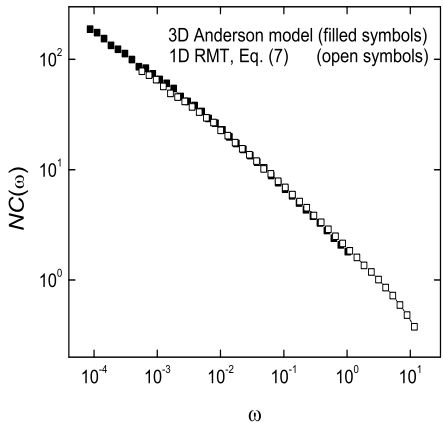}
\caption{A comparison of eigenfunction correlation functions for
3D Anderson model with $W=30$, $L=20$ and the truncated critical
RMT Eq.(7) with $b=0.42$, $B=5$, $\eta=1/3$ and $N=2000$. The IPR
takes values of $P_{2}=0.28$ and $0.25$, respectively. The scale
of $\omega=|E-E'|$ is different in those two cases by
approximately a factor of 11.}
\end{figure}
Now let us consider Eq.(\ref{C-finger}) for $d\geq 1$ assuming that
all states are exponentially localized and thus $\langle
|H_{nm}|^{2}\rangle\propto {\rm exp}[-R/\xi]$. We also assume for
simplicity that $F_{\omega}(R)=N^{-1}\,\theta(R-R_{0})$, where
$R_{0}=2\xi \ln(\delta_{\xi}/\omega)$. Then one immediately
obtains from Eq.(\ref{C-finger}) that due to the phase volume
factor $R^{d-1}$ the correlation of exponentially localized
eigenfunctions depends crucially on the dimensionality of space.
Namely, for $d=1$ the ideal insulator  limit Eq.(\ref{st-loc-l})
is reached for sufficiently small $\xi$ (see Fig.1), while for
$d>1$ and $\omega\ll \delta_{\xi}$ the correlation function
acquires a logarithmic in $\omega$ enhancement factor:
\begin{equation}
\label{ench} NC(\omega)\sim
\xi^{d-d_{2}}\,\ln^{d-1}\left(\frac{\delta_{\xi}}{\omega}\right).
\end{equation}
The physics behind this result is similar to the one which leads
to the selebrated Mott's law \cite{Mott,Ber} $\sigma(\omega)\propto 
\omega^{2}\;\ln^{d+1}(\delta_{\xi}/\omega)$ 
for the ac conductivity $\sigma(\omega)$ at $\omega\ll \delta_{\xi}$.
The difference is that the contribution to conductivity from the resonance states
with the distance $R$ between the points of maximal amplitude is proportional to the
square of the dipole moment $d^{2}\propto R^{2}$, so that the phase volume   
factor $R^{d-1}$ gets multiplied by $R^{2}$ resulting in emergence of the logarithmic 
factor $\ln^{d+1}(\delta_{\xi}/\omega)$ instead of $\ln^{d-1}(\delta_{\xi}/\omega)$
in our case.

Below we obtain this result for the {\it truncated critical} RMT
defined by Eq.(\ref{tran}).
The phase volume factor $R^{d-1}$  can be formally
taken into account in the random matrix formalism Eq.(\ref{k}) if
one assumes the following relationship between the $d$-dimensional
vector ${\bf R}$ and the difference of matrix indices $n-m$:
\begin{equation}
\label{repl} d(n-m)\Rightarrow \Omega_{d}\,R^{d-1}\,dR,\;\;\;
|n-m|\Rightarrow \frac{\Omega_{d}}{d}\,R^{d},
\end{equation}
where $\Omega_{d}$ is the total solid angle in the $d$-dimensional
space.

In particular Eq.(\ref{repl}) suggests that for exponential
localization the correct truncating factor in Eq.(\ref{tran}) has
the form:
\begin{equation}
\label{tr-fac} e^{-R/\xi}\Rightarrow {\rm
exp}\left[-\left(\frac{|n-m|}{B}\right)^{1/d}\right].
\end{equation}
This sets the exponent $\eta$ in Eq.(\ref{tran}) equal to:
\begin{equation}
\label{eta} \eta=\frac{1}{d}.
\end{equation}
Then Eq.(\ref{k}) can be used which is convenient to rewrite in
the following form:
\begin{equation}
\label{ka} NC(\omega)\approx k(\omega)=-\int_{0}^{\infty}
f(y)\,\frac{dy}{\frac{d}{dn}(\ln \sigma^{2}(n))|_{n=n(y)}},
\end{equation}
where $n(y)$ is found from the equation
$$y=\frac{\omega^{2}}{2\sigma^{2}(n(y))},$$ and
\begin{equation}
\label{f} f(y)=\left(\frac{2}{\sqrt{y}}+\frac{1}{y\sqrt{y}}
\right)\,e^{-y}\int_{0}^{\sqrt{y}}e^{t^{2}}\,dt
-\frac{1}{y}=\left\{\begin{matrix}\frac{4}{3}, & y\ll 1 \cr
\frac{1}{y^{2}}, & y\gg 1
\end{matrix}\right.
\end{equation}
For the truncated critical RMT Eq.(\ref{tran}) with $\eta=1/d$ one
finds: $$-\frac{1}{\frac{d}{dn}(\ln \sigma^{2}(n))}=
\frac{n}{2+\frac{1}{d}\,\left(\frac{n}{B}\right)^{1/d}},$$
where
\begin{equation}
\label{n-y}
n(y)=\left\{\begin{matrix}B\,\ln^{d}\left(\frac{2b^{2}y}{\omega^{2}B^{2}}
\right), &
\frac{\omega}{\sqrt{y}}\ll\delta_{\xi}\sim\frac{b}{B}\cr
\sqrt{\frac{2b^{2}y}{\omega^{2}}}, &
\frac{\omega}{\sqrt{y}}\gg\delta_{\xi}
\end{matrix} \right.
\end{equation}

 The integral in Eq.(\ref{ka}) is well convergent and
thus mainly contributed by $y\sim 1$. This makes it possible to
obtain a simple analytical expression for $NC(\omega)$:
\begin{equation}
\label{c-ffin} NC(\omega)\approx \left\{\begin{matrix}
c_{d}B\,\ln^{d-1}\left(\frac{\delta_{\xi}}{\omega} \right), &
\omega\ll\delta_{\xi}\sim b/B \cr
c_{0}\,\left(\frac{b}{\omega}\right), & E_{0}\sim b\gg\omega\gg
\delta_{\xi}
\end{matrix}\right.
\end{equation}
where $$ c_{d}=2^{d-1}d\,\int_{0}^{\infty}f(y)dy=2^{d}d.$$ $$
c_{0}=\int_{0}^{\infty}f(y)\,\sqrt{\frac{y}{2}}=\left(\frac{\pi}{2}
\right)^{\frac{3}{2}}\approx 1.97.$$ The first line of
Eq.(\ref{c-ffin})is consistent with Eq.(\ref{ench}) in which
$d_{2}/d\sim b\ll 1$ and $\xi\sim B^{1/d}$. The power-law behavior
in the second line of Eq.(\ref{c-ffin}) is a remnant of the
critical behavior Eq.(\ref{ChAnz}). It exists only for
considerably large $B\gg 1$ where $\delta_{\xi}\ll E_{0}=c_{0}b$, i.e. only in the 
{\it multifractal} insulator. For a very strong insulator with $\delta_{\xi}\gg 
E_{0}$, the localization radius is smaller than the minimal length scale $\ell_{0}$
of the fractal texture. This is the region of an ordinary insulator where
the entire  localization volume is more or less homogeneously filled.

Thus the eigenfunction correlation function for the truncated
critical RMT  Eq.(\ref{tran})interpolates  between the behavior
given by Eq.(\ref{ench}) at $\omega\ll\delta_{\xi}$ (or
$L_{\omega}=(\rho\omega)^{-1/d}\gg \xi$) and the critical behavior
Eq.(\ref{ChAnz}) which is valid for $\ell_{0}\ll L_{\omega}\ll \xi$. At
$L_{\omega}\sim \xi$ both asymptotic forms are apparently matching
with each other.

The physical picture that leads to such a behavior is the
following. There are two distinct regions $L_{\omega}<\xi$ and
$L_{\omega}>\xi$ where physics of eigenfunction correlations is
entirely different. In the case $L_{\omega}<\xi$ the characteristic
length $L_{\omega}$ has a meaning of the period of beating in the
overlap of two fractal eigenfunctions inside the localization
volume. The regions where two fractal supports match well with
each other alternate with the regions with a strong mismatch
between them, very much like in the case of two grids with
slightly different periods. The regions of strong overlap make the
main contribution $$C_{m}\sim P_{2}\sim
(L_{\omega}/\ell_{0})^{d-d_{2}}/\xi^{d}$$ to the eigenfunction
correlation function $C(\omega)$ which is of the order of the IPR
of a multifractal metal (see next Section for more details) with
the system size equal to the localization radius $\xi$ and the
correlation length equal to the size $L_{\omega}$ of the well
overlapping regions. To obtain the correlation function
$C(\omega)$ one has to multiply $C_{m}(\omega)$ by the probability
for the entire localization volumes to overlap. This probability
is $\xi^{d}/L^{d}$, as for $\omega>\delta_{\xi}$ there is no
correlations in the positions of the localization volumes. Thus 
we obtain the critical 
power-law Eq.(\ref{ChAnz}): $NC(\omega)\sim
(L_{\omega}/\ell_{0})^{d-d_{2}}=(E_{0}/\omega)^{1-d_{2}/d}$.

For $L_{\omega}>\xi$ physics of eigenfunction correlations changes
drastically. Now localization volumes are statistically repelling
each other and the overlap is only due to the tails. In this
region the length scale $L_{\omega}$  loses its physical meaning
which  is taken over by the length scale $R_{0}$ given by
Eq.(\ref{R}).

The overall shape of $C(\omega)$ with the logarithmic enhancement
factor $\ln^{d-1}(\delta_{\xi}/\omega)$ obtained within the
Truncated Critical RMT describes the numerical results on the 3D
Anderson model very well (see Fig.11). The absence of this factor
at $d=1$ explains the qualitative difference between the case
$d=1$ (see Fig.8) and $d=3$ (see Fig.9). This difference is
essentially due to a competition between two effects (i) repulsion
of centers of localization and (ii) resonance enhancement of
overlap by tails. The first effect tends to decrease the
probability of the overlap of localization volumes. The second
effect increases the eigenfunction overlap by means of tails. In
the 1D case these two effects compensate each other and the result
is the same as one would obtain for completely uncorrelated
positions of localization volumes and the typical exponentially
decreasing tails. In higher dimensions the enhancement of overlap
in the tail region prevails because of the increased volume of
those regions.

Concluding this subsection we claim that the truncated critical
RMT provides an excellent description of the 3D Anderson insulator
both in the strong localization region (see Fig.9) and in the
region of {\it multifractal insulator} where the localization
radius is large and the corresponding scale $\delta_{\xi}$ is
small compared with the upper cutoff $E_{0}$ of multifractal
correlations. Because of the limited size of the 3D lattice this
latter region is out of reach for numerical simulations on the 3D
Anderson model, and the random matrix theory is the only
mathematical model which properly describes physics of the
multifractal insulator.

\subsection{Super-critical PLBRM}
Note that there is another RMT  Eq.(\ref{alpha-ins}) suggested in
Ref. \cite{MFSeil} as a candidate to describe eigenfunction
correlations in the multifractal insulator. Below we show that
this {\it super-critical} PLBRM is principally flawed, as it
corresponds to a power-law localization which is not the case in
the 3D Anderson model.

This can be best demonstrated by Eq.(\ref{PT}) in which
$|H_{nm}|\Rightarrow\sigma(R)=(b/R)^{\alpha}$ with $\alpha>1$.
Accordingly, the typical scale for the repulsion of centers of
localization is:
\begin{equation}
\label{R-w} R_{\omega}= \frac{b}{\omega^{1/\alpha}}.
\end{equation}
In Fig.12 we plot the results of numerical calculation of the PDF
$F_{\omega}(R)$ for the super-critical PLBRM Eq.(\ref{alpha-ins}).
The characteristic scale $R_{\omega}$ where $F_{\omega}(R)$
reaches its maximum, is well seen in this plot.
\begin{figure}
\vspace{1cm}
\includegraphics[width=7cm, height=7cm]{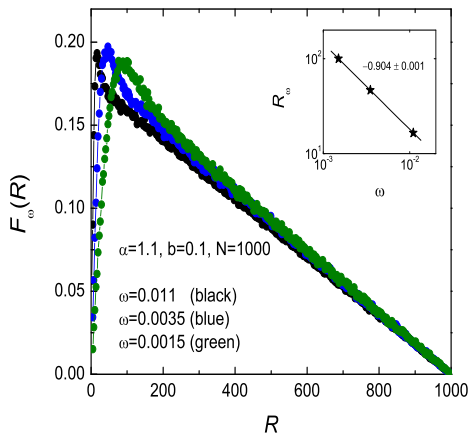}
\caption{The correlation function of centers of localization
$F_{\omega}(R)$ for the super-critical PLBRM Eq.(\ref{alpha-ins})
with $\alpha=1.1$, $b=0.1$, $N=1000$ for $\omega=0.011$ (black),
$0.0035$ (blue), $0.0015$ (green). The insert shows the
$\omega$-dependence of $R_{\omega}$ where $F_{\omega}(R)$ reaches
its maximum. The finite slope of $F_{\omega}(R)$ for $R\gg
R_{\omega}$ is a finite size effect which was neglected in
Eq.(\ref{F-w}).}
\end{figure}

The analytical treatment based on  Eq.(\ref{ka}) yields for this
model:
\begin{equation}
\label{C-mir}
NC(\omega)=\left(\frac{E_{\alpha}}{\omega}\right)^{\frac{1}{\alpha}}\sim
R_{\omega},
\end{equation}
where $E_{\alpha}=(c_{\alpha}b)^{\alpha}$, and
$$c_{\alpha}=\frac{\pi^{3/2}}{2^{1+\frac{1}{2\alpha}}}\,\frac{1}
{\Gamma(\frac{3}{2}-\frac{1}{2\alpha})}.$$ As well as the entire
approach based on Eq.(\ref{k}), the above results are valid when
$b$ is the smallest relevant parameter. In the problem of PLBRM
with $\alpha$ close to 1, there is a competition between the small
parameters  $|1-\alpha|$ and $b$, so that the validity of
Eqs.(\ref{R-w}),(\ref{C-mir}) requires also $\alpha-1\gg b$.

We see that in the infinite system $N\rightarrow\infty$ the
power-law Eq.(\ref{C-mir}) in $C(\omega)$ is not restricted at
small $\omega$, and no energy scale similar to $\delta_{\xi}$
emerges. This can be explained only if we assume that the
localization length for $\alpha-1>>b$ is of order one. Then for
all energy separations $\omega\gg E_{\alpha}$ the repulsion core
$R_{\omega}\gg \xi$, and no qualitative change in the correlation
function occurs until $R_{\omega}$ hits the system size $N$. For
smaller $\omega$ the correlation function is almost a constant.
This quantitative analysis  is illustrated by Fig.13.
\begin{figure}
\vspace{1cm}
\includegraphics[width=6cm, height=5cm]{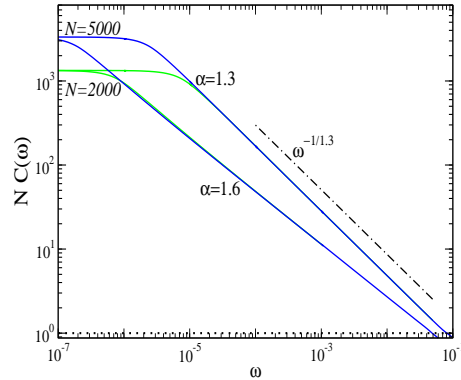}
\caption{Two-eigenfunction correlation for the super-critical
PLBRM Eq.(\ref{alpha-ins}) with $\alpha-1>>b$ calculated
analytically using Eqs.(\ref{C-K}-\ref{barO}). The power law
$C(\omega)\propto \omega^{-1/\alpha}$ is valid for all energy
separations corresponding to $1<R_{\omega}<N$. The onset of the
plateau moves to $\omega\rightarrow 0$ in the limit
$N\rightarrow\infty$. The ideal insulator limit is reached by
decreasing the slope with increasing $\alpha$.}
\end{figure}
\begin{figure}
\vspace{1cm}
\includegraphics[width=7cm, 
height=7cm]{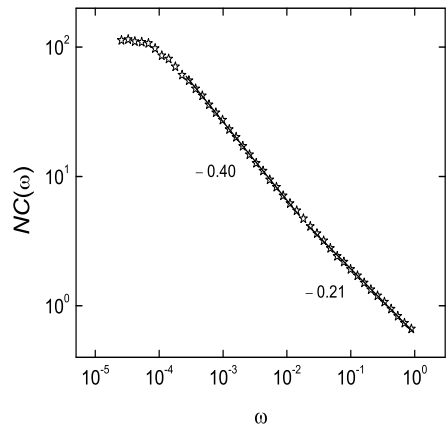}
\caption{Two-eigenfunction correlation for the super-critical
PLBRM Eq.(\ref{alpha-ins}). Numerical results for $\alpha-1 < b$
(this plot corresponds to $\alpha=1.15$, $b=0.45$, $N=5000$) show
that the exponent $\mu$ of the power-law $C(\omega)\propto
\omega^{-\mu}$ changes at $|E-E'|=\delta_{\xi}$ with the larger
value corresponding to smaller energy separations.}
\end{figure}

The region of $\alpha$ that could describe the multifractal
insulator with large $\xi\gg 1$ corresponds to $\alpha-1\ll b$. In
this case an energy scale similar to $\delta_{\xi}$ should appear.
It can be found from the condition
\begin{equation}
\label{d-xi}
R_{\omega}|_{\omega=\delta_{\xi}}=\xi(\alpha),\;\;\;\Rightarrow
\delta_{\xi}\approx \frac{E_{0}}{\xi}.
\end{equation}
At $|E-E'|=\delta_{\xi}$ the slope on the log-log plot of
$C(\omega)$ should change from the critical value at
$\delta_{\xi}<|E-E'|<E_{0}$ to a different (but constant)
$\alpha$-dependent value at $|E-E'|<\delta_{\xi}$. This change of
the slope is clearly seen in the numerical simulations on the
super-critical PLBRM presented in Fig.14. It appears that in all
cases studied  the slope $\mu$ at $|E-E'|<\delta_{\xi}$ is {\it
larger} than that at $|E-E'|>\delta_{\xi}$. This is in a clear
contradiction with the results (see Fig.9) obtained in the 3D
Anderson insulator.

An important conclusion we can draw from the above analysis of the
super-critical PLBRM is that the correlation function $C(\omega)$
for $\omega<\delta_{\xi}$ is the power-law in this model. This can
be traced back to the power-law character of localization in the
super-critical PLBRM which is not the case in the disordered
lattice models (such as the 3D Anderson model) with short-range
hopping integrals. This is the reason why the super-critical PLBRM
is not suitable to describe the insulating phase of the 3D
Anderson model.

\section{Search for random-matrix model for a multifractal metal}
\subsection{Anti-truncated critical RMT}
Surprisingly, the natural counterpart to the truncated critical
RMT Eq.(\ref{tran}) which is defined by Eq.(\ref{anti-trank})
("anti-truncated" critical RMT) does not describe extended states
in the multifractal metal. The reason is that this model possesses
{\it two} low-frequency system-size independent energy scales
instead of the single scale $\delta_{\xi}$ which is associated
with the size $\xi$ of a multifractal cell  in Fig.2c. In order to
see this we analyze the analytical formulae
Eqs.(\ref{C-K}-\ref{barO}) with the variance defined by
Eq.(\ref{anti-trank}).

To this end we expand the summand of Eq.(\ref{k}) in
$1/\bar{\omega} \ll 1$ to arrive at the formula similar to
Eq.(\ref{hvost}) but with the upper limit of integration equal to
the correlation radius $\xi$:
\begin{equation}
\label{kk} NC(\omega)=
\frac{2}{\omega^{2}}\int_{\sigma(n)<\omega}^{\xi}\sigma^{2}(n)\,dn.
\end{equation}
where $\xi=B/2b$ according to Eq.(\ref{xi}).

Substituting Eq.(\ref{anti-trank}) for $\sigma(n)$ we arrive at:
\begin{equation}
\label{kkk} NC(\omega)=
\frac{b^{2}}{\omega^{2}}\int_{b/\omega}^{\xi}\left[\frac{1}{n^{2}}+\frac{2}{B^{2}}
\right]\approx \frac{b}{\omega}+ 2\xi
\frac{b^{2}}{\omega^{2}B^{2}}.
\end{equation}
Eq.(\ref{kkk}) is valid at $\omega>b/\xi$ when the upper limit of
integration is larger than the lower limit. This sets the energy
scale $\omega_{1}=b/\xi = 2b^{2}/B$.

Another scale $\omega_{2}=1/(2b\xi)$ gives the cross-over scale
that separates the critical $1/\omega$ behavior and the
$1/\omega^{2}$ behavior that takes place for indermediate
frequencies $\omega_{1}<\omega<\omega_{2}$. While the scale
 $\omega_{1}=b/\xi$ (similar to the scale $\delta_{\xi}=b/B$ in a
 1D insulator)
determines the onset of the low-frequency plateau, the second
relevant scale $\omega_{2}$ that appears in the model
Eq.(\ref{anti-trank}) seems to have no physical meaning. 
Indeed,
the existence of this scale leads to a characteristic form of the
correlator $C(\omega)$ which log-log plot has a significant slope
increase just before it drops to zero at the plateau (see Figs.15,
16).
\begin{figure}
\vspace{1cm}
\includegraphics[width=7cm, height=7cm]{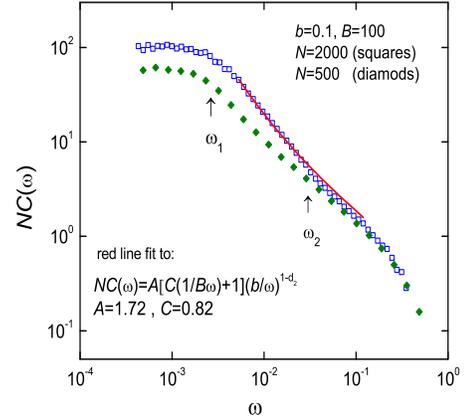}
\caption{Numerics on the anti-truncated critical RMT of the
orthogonal symmetry class. The predicted analytically
non-monotonous behavior of log-log slope controlled by the energy
scale $\omega_{2}$  is well seen both for small and large system
sizes. The solid line is a fit according to Eq.(\ref{kkk}).}
\end{figure}

We did not find the behavior of such type in the 3D Anderson metal
(see Fig.17). The plot in Fig.17 clearly shows a saturation
\cite{A-m} at $\omega<\omega_{1}$:
\begin{equation}
\label{satur} C(\omega)\approx \frac{1}{3}
P_{2}=A_{m}\,\frac{\xi^{d-d_{2}}}{3L^{d}},\;\;\;\;A_{m}\approx
0.5.
\end{equation}
However, there is no evidence of a maximum in the slope just above
the onset of the plateau.

\begin{figure}
\vspace{1cm}
\includegraphics[width=5cm, height=5cm]{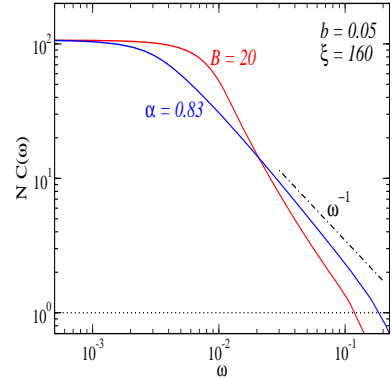}
\caption{Eigenfunction correlation for the anti-truncated critical
RMT Eq.(\ref{tran}) (red curve) and for the sub-critical PLBRM
Eq.(\ref{alpha-ins}) (blue curve) of unitary symmetry class
computed analytically from Eqs.(\ref{C-K}-\ref{barO}).  For the
anti-truncated critical RMT the non-monotonous behavior of log-log
slope is similar to the one in Fig.15.}
\end{figure}
\begin{figure}
\vspace{1cm}
\includegraphics[width=8cm, height=8cm]{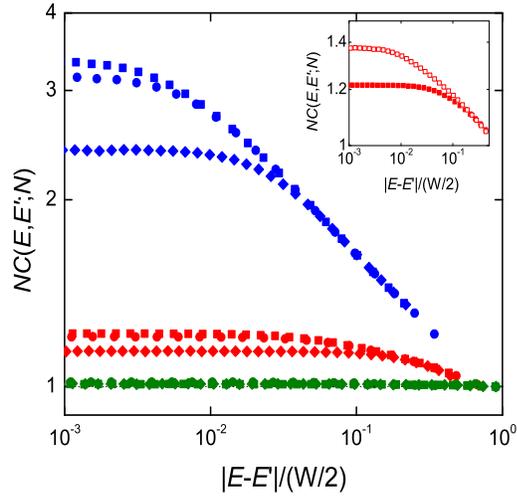}
\caption{Eigenfunction correlation in the 3D Anderson model:
extended states. The disorder strength is $W=2$ (green), $W=5$
(red), $W=10$ (blue). The system size is $L=8$ (diamonds) $L=16$
(circles) and $L=20$ (squares). The correlation length is
estimated from the inverse participation ratio as follows
$\xi=(P_{2}N/A_{m})^{1/(d-d_{2})}$ ($A_{m}=0.5$, $d_{2}=1.3$) and
is equal to: $\xi=5.8$, $3.2$, $2.8$ at $W=10$, $5$, $2$,
respectively. In the insert we compare results for $W=5$, $L=20$
for the periodic (red filled squares) and the hard-wall (red open
squares) boundary conditions. For hard-wall boundary conditions
the correlation function looks "more critical", as the critical
point $W_{c}=15.2$ in this case is closer to $W=5$.}
\end{figure}
\subsection{Sub-critical PLRBM}
Now we consider the sub-critical PLBRM ensemble defined by
Eq.(\ref{alpha-ins}) with $1/2<\alpha<1$. In this case analytical
arguments similar to Eqs.(\ref{kk},\ref{kkk}) predict only one
relevant energy scale $\omega_{1}=b/\xi^{\alpha}$ such that for
$\omega< \omega_{1}$ the correlation function $C(\omega)$ is
constant and for $\omega>\omega_{1}$ (but $\omega<E_{0}\sim
b^{\alpha}$) it is a pure power law $C(\omega)\propto
\omega^{-\frac{1}{\alpha}}$. Thus the sub-critical PLBRM is free
from the drawback related with the unphysical second energy scale.
For comparison we plotted the analytical results for the
anti-truncated critical RMT and for the sub-critical PLBRM in
Fig.16. It is seen that the overall shape of the blue curve for
sub-critical PLBRM is much closer to the results of 3D Anderson
model of Fig.17.

Note however, that the power-law emerging in the analytical
results for the  sub-critical PLBRM has an exponent $1/\alpha$
which is {\it larger} than the critical exponent. Computer
simulations (see Fig.18a) on the sub-critical PLBRM with very
small $b$ confirm this analytical result as $N\rightarrow\infty$
extrapolation and show that the slope increases with increasing
the system size N. However, the slope of the corresponding curves
for the 3D Anderson model of Fig.17 is almost independent of the
system size and is  equal or {\it smaller} than the critical slope.

\begin{figure}
\vspace{1cm}
\includegraphics[width=9.5cm, height=7cm]{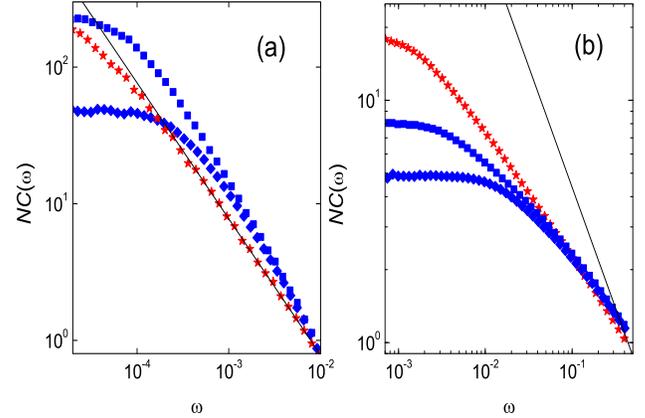}
\caption{Eigenfunction correlations in the sub-critical PLBRM of
the orthogonal symmetry class. Left panel (a): the small-$b$ limit
$b=0.001$, $\alpha=0.8$ at $N=1000$ (squares) and $N=200$
(diamonds). The critical correlations for $b=0.001$, $\alpha=1$
and $N=1000$ are shown by  stars. The solid line corresponds to
the power-law $\omega^{-1}$. The slope of the sub-critical curve
is larger than 1 (correlations are short-range in the energy
space). This case is relevant for the metallic phase of the
Anderson model in very high dimensions. Right panel (b): the case
$b=0.4$, $\alpha=0.95$ at $N=1000$ (squares) and $N=200$ (diamonds)
is relevant for the Anderson model in $d=3$ ("multifractal"
metal). The slope is less than the critical (which in turn is less
than 1) and is almost size-independent. Correlations are
long-range in the energy space.}
\end{figure}

The reason for the discrepancy is that the analytical result for
the slope $\mu=1/\alpha$ corresponds to the limit $b\rightarrow
0$. At a finite $b$ the slope decreases with increasing $b$ and at
a sufficiently large $b$ may become smaller than the critical one.
It is reasonable to assume that at small $1-\alpha$ and $b$ this
happens at $b\sim (1-\alpha)$. The relevance of the parameter
$b/(1-\alpha)$ is also seen from the expression for the
correlation length $\xi$
\begin{equation}
\label{xi-anal} \xi\sim {\rm
exp}\left[\frac{a_{\alpha}}{1-\alpha}\,
\right],\;\;\;\;a_{\alpha}=\ln\left(\frac{1-\alpha}{b^{\alpha}}\right)+{\rm
const }.
\end{equation}
which was found (up to a constant of order one) analytically from Eq.(\ref{xi}).
This expression is apparently meaningless for $b^{\alpha}\gg
(1-\alpha)$ where $c_{\alpha}$ may become negative.

Numerical simulations on the sub-critical PLBRM with $1-\alpha <
b$ (e.g. for $\alpha=0.95$ and $b=0.4$ relevant for the 3D
Anderson model) show (see Fig.18b) that the log-log slope of
$C(\omega)$  is somewhat smaller than the critical one and is almost
independent of the matrix size $N$. Thus the sub-critical PLBRM
shows exactly the same  character of eigenfunction correlations as
in the 3D Anderson metal (see Fig.17).

Two parameters of the sub-critical PLBRM allow to simulate the
effect of the finite correlation length (choice of $\alpha<1$) and
the dimensionality of space (choice of $b$). Note in this
connection that for the disorder strength $W$ significantly
smaller than the critical value $W_{c}$, not only $\alpha$ but
also $b$ is $W$-dependent. The point is that in the 3D Anderson
model the variance of the on-site energies fluctuations is
proportional to $W^{2}$, while the off-diagonal hopping integral
is equal to 1. This implies that the ratio of a typical
off-diagonal to a typical diagonal elements controlled in
Eq.(\ref{alpha-ins}) by the parameter $b$ should scale like $1/W$.
As the log-log slope of $C(\omega)$ decreases with increasing $b$,
moving away from the Anderson transition into the {\it metallic}
phase $W<W_{c}$ has an effect of {\it decreasing} the slope. On
the insulator side of the transition the situation is opposite and
one should expect an {\it increase} of the slope (for
$\delta_{\xi}<\omega<E_{0}$) with increasing $W$. Fig.9 shows that
it is apparently the case.

Another relevant note is that for the Anderson model in {\it
higher dimensions} the correlation dimension $d_{2}$ decreases.
This can be modeled by a decreasing  parameter $b$. Then the
analogy with the sub-critical PLBRM suggests that for sufficiently
high dimensions $d>d_{c}$ the behavior in the $d$-dimensional
Anderson model  should become similar to the one in Fig.18a.
Namely, the exponent $\mu$ in Eq.(\ref{d-mu}) may become {\it larger
than 1}. This changes qualitatively the eigenfunction
correlations, as they become effectively short-range in the energy
space. In particular, the return probability \cite{CKL} which is 
proportional to the Fourier transform of $C(\omega)$ behaves
in the time interval $\hbar/E_{0}\ll t\ll \hbar/\delta_{\xi}$
as $P(t)\propto t^{-(1-\mu)}$ for $\mu<1$ and is a constant for
$\mu>1$.

We believe that this qualitative change in the
eigenfunction statistics (if confirmed for a $d$-dimensional
Anderson model with $d>d_{c}$) should lead to dramatic physical
consequences marking a transition to a new metallic state.
\section{Conclusion}
In conclusion we list the main results obtained above. The most
important of them is the persistence -- beyond the point of
localization transition-- of the critical power-law in the
dependence of the eigenfunction correlation function $C(\omega)$ on the
energy separation $\omega$ and the related enhancement of
$C(\omega)$ at $\delta_{\xi}\ll\omega\ll E_{0}$, where
$\delta_{\xi}$ is the mean level spacing in the
localization/correlation volume and $E_{0}$ is the upper energy
cut-off of multifractality. This enhancement leads to an
enhancement of matrix elements of local electron interaction which
may result in, e.g. an enhancement of the superconducting
transition temperature in the vicinity of the Anderson
localization transition \cite{FIKY}. Another important observation
is that the enhancement of correlations at $\omega<E_{0}$ is always accompanied by
the depression at $\omega> E_{0}$, both phenomena being the
consequences of the stratification of the coordinate space into
mutually avoiding supports of the fractal structure with well
overlapping eigenfunctions living on each of them. An independent
-- but also important-- phenomenon is the logarithmic enhancement
of $C(\omega)$ in the 2D and 3D Anderson insulator at
$\omega<\delta_{\xi}$ (and the
absence of such enhancement in the quasi-1D disordered wire). 
It is a result of a competition of two
simultaneous phenomena: the repulsion of centers of localization
and the resonance enhancement of the eigenfunction overlap by
tails. Both phenomena are studied quantitatively within the
Truncated Critical Random Matrix model which is suggested as a
universal tool to describe the localized eigenfunctions with a
multifractal texture. We also show that the sub-critical Power-Law
Banded Random Matrix Ensemble suggested in Ref.\cite{MFSeil}
describes the multifractal metal reasonably well. From the
analytical solution for this RMT we conclude that a critical
dimensionality $d_{c}$ may exist above which the $d$-dimensional
Anderson model has an unusual metal phase characterized by an effectively
short-range correlation function $C(\omega)\propto \omega^{-\mu}$
with $\mu >1$.

{\it Acknowledgement.}-- The authors are grateful to B.L.Altshuler,
M.V.Feigel'man, A.Silva and V.I.Yudson for stimulating discussions and
especially to O.Yevtushenko for a collaboration at an earlier
stage of this work and for a help in preparing figures.
E.C. thanks the FEDER and the Spanish DGI for financial
support through Project No. FIS2004-03117.


\begin{thebibliography}{99}
\bibitem{A-m} The constant $A_{m}$ (and the corresponding constant
$A_{i}$) for insulator were found from an independent numerical
study of an exponential tail
$|\Psi(r)|^{2}-|\Psi(\infty)|^{2}\propto e^{-r/\xi}$ of the wave
function, with $r$ counted from the maximum of the wave function
amplitude.
\end{thebibliography}
\end{document}